\documentclass[twocolumn]{aastex631}

\newcommand{\Msun}{\,{\rm M_\odot}}

\newcommand{\Mblack}{M_\bullet}

\begin{document}

\title{Extreme Tidal Stripping May Explain the Overmassive Black Hole in Leo I: a Proof of Concept}

\correspondingauthor{Fabio Pacucci}
\email{fabio.pacucci@cfa.harvard.edu}

\author[0000-0001-9879-7780]{Fabio Pacucci}
\affiliation{Center for Astrophysics $\vert$ Harvard \& Smithsonian, Cambridge, MA 02138, USA}
\affiliation{Black Hole Initiative, Harvard University, Cambridge, MA 02138, USA}

\author[0000-0001-7899-7195]{Yueying Ni}
\affiliation{Center for Astrophysics $\vert$ Harvard \& Smithsonian, Cambridge, MA 02138, USA}

\author[0000-0003-4330-287X]{Abraham Loeb}
\affiliation{Center for Astrophysics $\vert$ Harvard \& Smithsonian, Cambridge, MA 02138, USA}
\affiliation{Black Hole Initiative, Harvard University, Cambridge, MA 02138, USA}



\begin{abstract}
A recent study found dynamical evidence of a supermassive black hole of $\sim 3 \times 10^{6} \Msun$ at the center of Leo~I, the most distant dwarf spheroidal galaxy of the Milky Way. This black hole, comparable in mass to the Milky Way's Sgr A*, places the system $>2$ orders of magnitude above the standard $\Mblack-M_{\star}$ relation. We investigate the possibility, from a dynamical standpoint, that Leo~I's stellar system was originally much more massive and, thus, closer to the relation. Extreme tidal disruption from one or two close passages within the Milky Way's virial radius could have removed most of its stellar mass. A simple analytical model suggests that the progenitor of Leo~I could have experienced a mass loss in the range of $32\% - 57\%$ from a single pericenter passage, depending on the stellar velocity dispersion estimate. This mass loss percentage increases to the range $66\% - 78\%$ if the pericenter occurs at the minimum distance allowed by current orbital reconstructions. Detailed N-body simulations show that the mass loss could reach $\sim 90\%$ with up to two passages, again with pericenter distances compatible with the minimum value allowed by Gaia data. Despite very significant uncertainties in the properties of Leo~I, we reproduce its current position and velocity dispersion, as well as the final stellar mass enclosed in $1$ kpc ($\sim 5\times 10^6 \Msun$) within a factor $< 2$. The most recent tidal stream is directed along our line of sight toward Leo I, making it difficult to detect. Evidence from this extreme tidal disruption event could be present in current Gaia data in the form of extended tidal streams.
\end{abstract}


\section{Introduction} 
\label{sec:intro}
Leo~I is the most distant dwarf spheroidal (dSph) galaxy associated with the Milky Way (MW) system. 
Located at a distance of $255$ kpc, with a total mass of $(8.1\pm 2.0) \times 10^7 \Msun$, and a high radial velocity of $ 287 \, \rm km \, s^{-1}$ \citep{Caputo_1999, Bellazzini_2004, Mateo_2008}, this dwarf galaxy is of great interest for dynamical studies of the neighborhood of the MW.

Leo~I recently became the subject of renewed interest because of what it may harbor at its center. The peculiar kinematics of stars at the core of Leo~I suggested that it could host a supermassive black hole (SMBH), with a mass of $(3.3 \pm 2) \times 10^6 \Msun$ \citep{Bustamante_2021}. Relying on a detailed dynamical analysis, the study argues that the no-black-hole case is excluded at $>95\%$ significance level in all the models considered.
Remarkably, the mass of this object would be similar to that of the SMBH at the center of the MW \citep{EHT_SgrA}, which was identified and localized by analyzing the stellar orbits in its vicinity \citep{Ghez_1998, Genzel_2000, Genzel_2003, Ghez_2005, Ghez_2008, Genzel_2010}. If the presence of this SMBH is confirmed, this places the Leo~I system more than two orders of magnitude above standard scaling relations between the mass of central SMBHs and the stellar mass of their hosts; the question of what process may have created such an overmassive black hole is then crucial.
The existence of this SMBH would also open up the exciting possibility of observing such an object electromagnetically \citep{Pacucci_2022_Leo}.

Prior to the dynamical detection of the SMBH, Leo~I already offered a vital laboratory to investigate the MW's and its neighbors' dynamical properties. For example, Leo~I is a valuable tool for estimating the total mass of the MW Galaxy \citep{Kochanek_1996, BK_2013}. 
Furthermore, investigations of the impact of MW tides on the dynamics and star formation history of satellites have implications for the distribution of dark matter on smaller scales \citep{Sohn_2007, BK_2013, Dooley_2016, Hausamann_2019, Wang_2023}.
In this context, Leo~I is a remarkable case due to its considerable galactocentric distance, which makes it less likely to be affected by tidal forces. Comparing its dynamics with those of dSph galaxies closer to the MW informs us of our Galaxy's impact on their evolution.

Historically, one of the most crucial unknowns regarding Leo~I was its orbit and its pericenter distance with the MW. In particular, is Leo~I bound to the MW? Early studies suggested that this dSph galaxy was on a hyperbolic orbit bound to M31 \citep{Byrd_1994}. However, \cite{Zaritsky_1989} had already argued that Leo~I was bound to the MW and used its orbit to estimate the mass of our Galaxy. More recent studies (see, e.g., \citealt{Mateo_2008, Sohn_2013}) suggest a closed orbit; \cite{Pace_2022} used Gaia EDR3 precision astrometry to constrain its orbital parameters. According to these latest data, Leo~I is on a closed, quite eccentric orbit with pericenter $47.5^{+30.9}_{-24.0}$ kpc and eccentricity $0.79^{+0.10}_{-0.09}$.

Its average value of the pericenter, $R_p \sim 50$ kpc, places Leo~I well within the sphere of influence of the MW, which has a virial radius of $\sim 200$ kpc \citep{Dehnen_2006}.
Whether Leo~I underwent significant tidal disruption during its close encounter with the MW is an investigation worth pursuing, as this process may have created a system with a very overmassive black hole at its center.

Some early studies already suggested that Leo~I was likely tidally disrupted on at least one perigalactic passage, and two at most, based on evidence of a flat velocity dispersion profile \citep{Mateo_2008} and a somewhat asymmetric radial velocity distribution of Leo~I stars, especially at large distances \citep{Sohn_2007}. 
Several studies have investigated the tidal stripping scenario for MW satellites with observational data \citep{Wang_2023}, large-scale simulations \citep{Chang_2013, Fattahi_2018, Hausamann_2019, Mazzarini_2020}, N-body simulations \citep{Sales_2010} theoretical models \citep{Kang_2008, Henriques_2010}. Interestingly, the tidal stripping efficiency depends strongly on the morphology of the satellite, with disk-like structures being the most disrupted \citep{Chang_2013}. 

A significant tidal disruption event for Leo~I would have caused tidal tails, which are, to date, unobserved. However, these tidal tails could coincide with our line of sight and thus be challenging to detect \citep{Mayer_2008}.

In this study, we develop a simple analytical model to estimate the effect of tidal disruption in Leo~I, and we support it with detailed N-body simulations of the MW + Leo~I system. 

\section{Methods}
\label{sec:methods}
In the following, we estimate the mass loss of a satellite (i.e., Leo~I's progenitor) due to tidal stripping.
First, we develop a simple analytical model to estimate the mass loss during a single pericenter passage. Then, we test this estimate via detailed N-body simulations in the multi-passage regime.

A note of clarification on the scope of this work is warranted. Significant uncertainties characterize Leo~I's system: its (putative) SMBH mass, its stellar mass, and its light-to-mass ratio are all estimated with uncertainties of up to $\sim 60\%$. Even its position on the $M_\bullet-M_\star$ plane is highly uncertain, given typical scatters of $\sim 0.5$ dex. With all these uncertainties, our work is intended as a proof of concept. Our goal is to show that Leo~I's system can undergo extreme tidal stripping and reproduce approximately its current physical properties and distance from the MW.

\subsection{Analytical Model}
\label{subsec:theory}

Satellite galaxies, orbiting inside the dark matter halo of their hosts, experience tidal forces from the massive host galaxy. This process gradually strips stars from the outer parts of the satellite; if the orbit is bound, and multiple pericenter passages are allowed, the entire satellite galaxy will eventually be disrupted within the host \citep{Galactic_Dynamics_2008}.

To estimate the mass loss during a single passage at the pericenter radius, $R_p$, we assume that the satellite, of mass $m$, is described by a singular isothermal sphere:
\begin{equation}
    \rho(r) = \frac{\sigma^2}{2\pi G r^2} \, ,
\end{equation}
where $G$ is the gravitational constant and $\sigma$ is the velocity dispersion, which is related to the circular velocity $v_c$ as $\sigma= v_c/\sqrt{2}$.
Note that in a singular isothermal sphere, the circular velocity is constant. This simple model is well suited for Leo~I because its dispersion profile is flat from its core to beyond its tidal radius, with mean velocity dispersion of $9.2 \pm 0.4 \, \rm km \, s^{-1}$ \citep{Mateo_2008}, or $11.76 \pm 0.66 \, \rm km \, s^{-1}$ \citep{Bustamante_2021}.

Fundamental to this calculation is the concept of tidal radius: the distance from the center of mass of the satellite where a star becomes gravitationally unbound (e.g., stripped). The star then becomes bound to the massive galaxy that the satellite is orbiting \citep{Read_2006}.
Our analytical model assumes that the tidal stripping occurs entirely at the pericenter during a single passage. Furthermore, we express the mass loss as a fractional change, i.e., $\Delta m/m$.

The tidal radius $R_t$ of Leo~I, at the pericenter passage of radius $R_p$, is estimated as follows \citep{vonHoerner_1957, King_1962, Spitzer_1987, Read_2006, Galactic_Dynamics_2008}:
\begin{equation}
    R_t = R_p \left[ \frac{m}{M(3+e)} \right]^{1/3} \, ,
\end{equation}
where $M$ is the mass of the MW (interior to the pericenter radius), and $e$ is the eccentricity of the satellite's orbit. This latter parameter is affected by large uncertainties \citep{Pace_2022}, but, given the exponent $1/3$, the quantities in the parenthesis need not be known with great accuracy \citep{Galactic_Dynamics_2008}. We then assume a conservative value of $e=0.5$.
From these expressions, we calculate $\Delta m = m - m(R_t)$, where the last term indicates the satellite's mass within its tidal radius (i.e., the mass of the satellite that is retained).

We obtain an estimate of the fractional mass loss, which is a function of the velocity dispersion of the satellite, its pericenter radius, and the masses of the satellite and its host: 
\begin{equation}
    \frac{\Delta m}{m} = 1 - \frac{2 \sigma^2}{G} R_p m^{-2/3} (3.5M)^{-1/3} \, .
\end{equation}

\subsection{N-body Simulations}
\label{subsec:sims}

As a proof of concept, we model the scenario of extreme tidal stripping with N-body simulations.
Our goal is to prove that extreme tidal stripping can reproduce, approximately, the observed properties of Leo~I. We note, however, that the requirements for this to occur are strict, such as the necessity of more than a single close pericenter passages. Given the significant uncertainties in the physical properties of Leo~I, our goal is to show that the extreme tidal stripping scenario is physically plausible. A detailed statistical analysis of the likelihood of this scenario involves a suite of N-body simulations and is beyond the scope of this work.

We use the public version of Arepo \citep{Arepo_2020} to conduct the N-body simulations in this study. The gravitational evolution of the system is modeled by the tree-based gravity solver, with the direct summation method enabled when less than $500$ particles are active at the tree-based timesteps \citep[see, e.g.,][for details of the algorithm]{Springel2001NewA....6...79S}. 
As our focus is to study the effect of tidal disruption of Leo~I, for which the gravitational effect makes the dominant contribution, we argue that a pure N-body simulation is sufficient for this proof-of-concept study.

In summary, we are interested in the following experiment:
\begin{enumerate}
    \item We start from the current 3D position of Leo~I and, in the MW potential, integrate its orbit back for a few Gyr.
    \item In this early time, we increase the total stellar mass $M_\star(r< 10 \, \mathrm{kpc}$) of Leo~I to $\sim 10^9 \Msun$ to place it on the local $\Mblack-M_\star$ relation \citep{Schutte_2019, Reines_Volonteri_2015}. In this ``ancient'' and ``augmented'' Leo~I, we place a SMBH with the mass estimated by \cite{Bustamante_2021}. We do not model the black hole mass growth and its radiative interaction with the environment within the N-body simulation.
    \item We then evolve the system in time to reproduce at $z=0$ the current properties of Leo~I. In this time frame, we study how its stellar and dark matter mass is affected by tidal stripping.
\end{enumerate}

We generate the initial conditions of isolated disk galaxies using \texttt{MakeDisk} \citep{Springel_DiMatteo_2005}. 
This code solves the Jeans equations for a quasi-equilibrium collisionless system of halo, disk, and bulge, with the a Maxwellian particle distribution function in the velocity space. 
We set up the isolated system of the MW using the parameters adopted in the AGORA Project \citep{AGORA_2014, Agora_2016}.
The dark matter halo of the MW has mass $M_{200} = 1.07 \times 10^{12} \Msun$, following an NFW \citep{Navarro_1997} profile with corresponding concentration parameter $c = 10$ and spin parameter $\lambda = 0.04$. 
The disk component of the MW has mass $M_d = 3.4 \times 10^{10} \Msun$, following an exponential profile with scale length $r_d = 3.4$ kpc and scale height $z_d = 0.1 \, r_d$. 
The stellar bulge has mass $M_b = 0.34 \times 10^{10} \Msun$ (i.e., bulge-to-disk mass ratio B/D = 0.1), following the Hernquist density profile \citep{Hernquist_1990}.

We also initialize the progenitor of Leo~I as an isolated disk galaxy with halo mass $M_{200} = 1.0 \times 10^{10} \Msun$ and total stellar mass $M_* = 1.2 \times 10^{9} \Msun$, which places the system exactly on top of the $\Mblack-M_\star$ relation \citep{Schutte_2019}.
We set the resolution of the N-body simulation with a dark matter mass of $M_{\rm dm} = 10^6 \Msun$ and a stellar disk/bulge mass of $M_{\rm *} = 10^5 \Msun$. 
The gravitational softening lengths are $\epsilon = 0.2$ kpc and $0.05$ kpc for dark matter and stars, respectively.

We use  \href{http://github.com/jobovy/galpy}{GalPy} \citep{GALPY_2015} to design the possible orbits of Leo~I that can have two pericenter passages over the past $8$ Gyr. These two passages are consistent with studies on its extended morphology and dynamics, suggesting that Leo~I has been tidally disrupted on at least one, but at most two, pericenter passages \citep{Sohn_2007}. 
In our N-body simulation, the orbit has minimum pericenter distances from the MW in the range of $15-20$ kpc and can recover the current galactocentric radial position of Leo~I. This orbit is consistent with Gaia EDR3-based studies on the orbits of MW dwarf spheroidal galaxies \citep{Pace_2022}.
With those constraints, we set the initial relative position and velocity of Leo~I progenitor as $(r_x, r_y, r_z) = (-104.9, -69.4, -76.1)$ kpc and $(v_x, v_y, v_z) = (146.8, 54.3, 131.6)$ $ \rm km \, s^{-1}$.

\section{Results}
\label{sec:results}
In this Section, we detail the results of our analytical model and N-body simulations.

\subsection{Theoretical Model: Single Pericenter Passage}
\begin{figure}%
    \centering
\includegraphics[angle=0,width=0.5\textwidth]{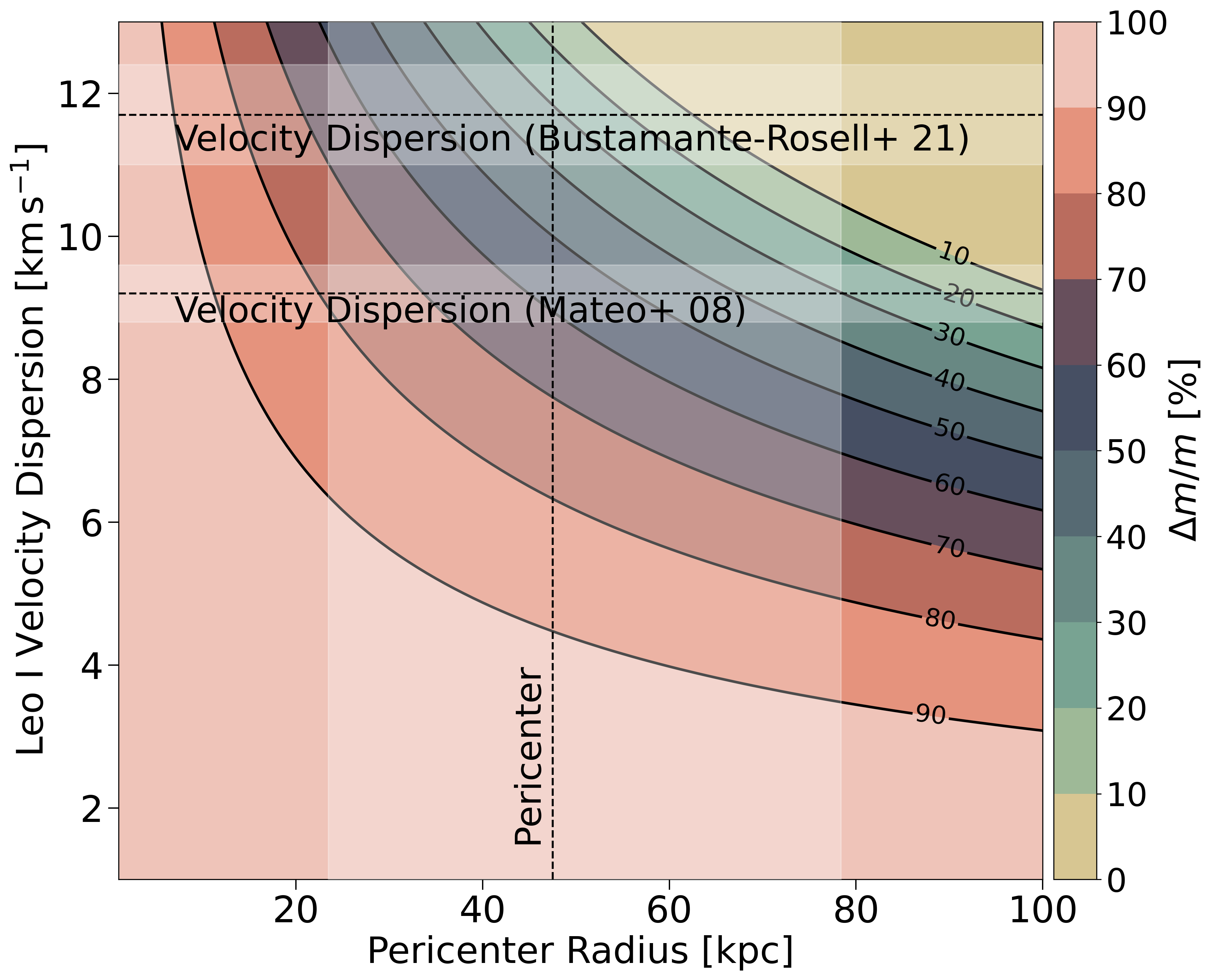}
    \caption{Fractional mass loss of the progenitor of Leo~I, calculated analytically for a single pericenter passage. The color scale shows the fractional mass loss as a function of the pericenter radius and the progenitor's stellar velocity dispersion. The horizontal and vertical bands show estimates of the pericenter and of the velocity dispersion of Leo~I (according to two studies), with $1\sigma$ confidence intervals.}
    \label{fig:theory}%
\end{figure}

Figure \ref{fig:theory} displays the fractional mass loss of the progenitor of Leo~I as a function of the pericenter radius and the stellar velocity dispersion. The horizontal and vertical bands show current estimates, with $1\sigma$ uncertainties, for the velocity dispersion, $9.2 \pm 0.4 \, \rm km \, s^{-1}$ \citep{Mateo_2008}, and the pericenter distance $47.5^{+30.9}_{-24.0}$ kpc \citep{Pace_2022}. Note that \cite{Bustamante_2021} estimate a higher value of the velocity dispersion, $11.76 \pm 0.66 \, \rm km \, s^{-1}$, due to crowding effects.

With the velocity dispersion estimated by \cite{Mateo_2008}, our model predicts that a single pericenter passage can strip $57\%$ of the initial mass of the progenitor. A closer pericenter passage would allow up to $78\%$ of mass loss, while a farther one would reduce the mass loss to $30\%$. Overall, depending on the pericenter, we estimate a mass loss of $57^{+21}_{-27} \%$ for a single passage.
In contrast, with the velocity dispersion estimated by \cite{Bustamante_2021}, our model predicts a mass loss of $32^{+34}_{-32} \%$ for a single passage.

\begin{figure}%
    \centering
\includegraphics[angle=0,width=0.5\textwidth]{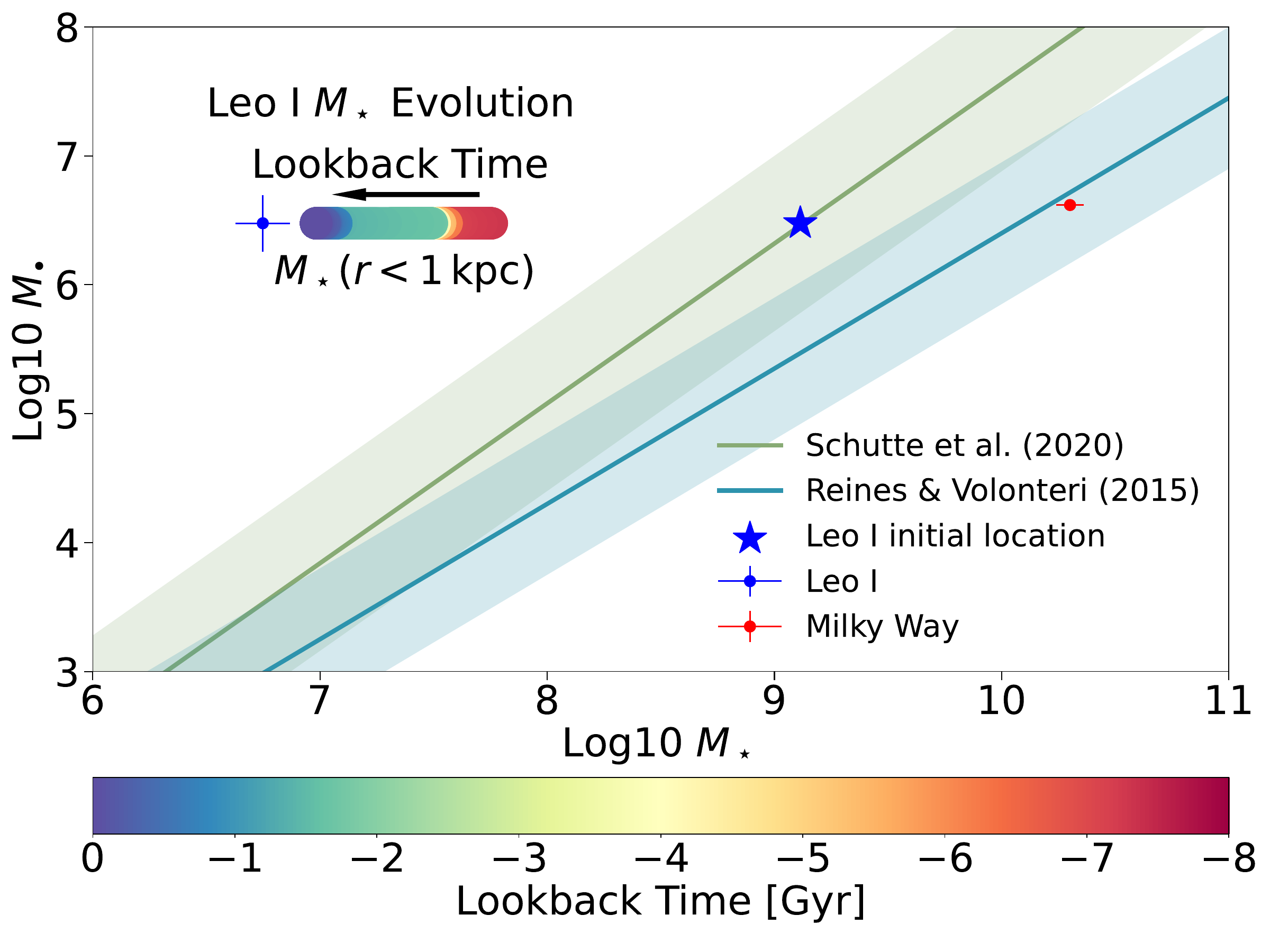}
    \caption{$M_\bullet-M_\star$ plane, displaying: (i) the current location of the MW and Leo~I (with their $1\sigma$ uncertainties in the mass estimates), (ii) the relations by \citealt{Schutte_2019} (appropriate for dwarf galaxies) and \citealt{Reines_Volonteri_2015}, (iii) the time evolution, in terms of the lookback time (see the color bar), of the stellar mass enclosed within $1$ kpc from our N-body simulations, and (iv) the initial location in the $M_\bullet-M_\star$ plane of our simulated Leo~I system, with the total stellar mass enclosed within $10$ kpc.}
    \label{fig:Mmstar_plot}%
\end{figure}

\begin{figure*}%
    \centering
    \includegraphics[angle=0,width=0.99\textwidth]{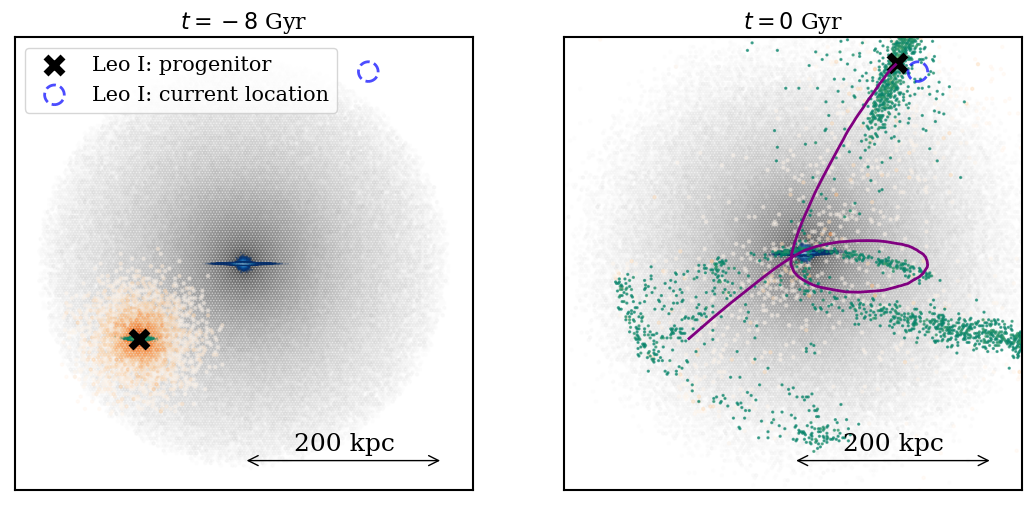}
    \caption{Visual representation of our N-body simulation, with the left and right panel showing the initial and final state of the simulation, respectively. Gray and blue colors give the projected density of dark matter and stellar components of the MW. Orange and green colors depict the projected density of the dark matter and stellar component of Leo~I's progenitor, displaying their dispersion across the MW halo caused by tidal disruption. The purple line traces the orbit of Leo~I's black hole across the 8 billion-year duration of the simulation. The panels are in an edge-on view of the MW disk.}
    \label{fig:image}%
\end{figure*}

\subsection{N-body Simulations: Multiple Pericenter Passages}

We run a series of N-body simulations to test this model, as described in Sec. \ref{subsec:sims}.
Figure \ref{fig:Mmstar_plot} shows the time evolution in the $M_\bullet-M_\star$ plane of the stellar mass enclosed within $1$ kpc in our simulations, while Figure \ref{fig:image} visually represents our reference run. The progenitor passes at the pericenter with the MW two times \citep{Sohn_2007}, with pericenter distances in the range of $15-20$ kpc, which is close to the lower-bound recently estimated with Gaia EDR3 data \citep{Pace_2022}. The reconstructed trajectory is entirely within the dark matter halo of the MW.

The time evolution track (Fig. \ref{fig:Mmstar_plot}, where the lookback time is indicated in the color bar) shows a stellar mass steadily decreasing over $8$ Gyr, while the mass of the SMBH is kept at the current value estimated by \cite{Bustamante_2021}. The progenitor of Leo~I starts as a somewhat extended galaxy with $\sim 10\%$ of the current stellar mass of the MW. After two pericenter passages, it reaches a final mass of $9 \times 10^6 \Msun$ within $1$ kpc. 
Note that \cite{Mateo_2008} estimate a stellar mass of $\sim (5.6 \pm 1.8)\times 10^6 \Msun$ within $1$ kpc, i.e., a total mass of $(8.1 \pm 2.0) \times 10^7 \Msun$, with a mass-to-light ratio of $(14.4\pm 5.8)$. 
Additionally, the final velocity dispersion estimated from our simulation is $11.3 \, \mathrm{km \, s^{-1}}$, which is higher than the value reported by \cite{Mateo_2008}, $9.2 \pm 0.4 \, \mathrm{km \, s^{-1}}$, but in agreement with the more recent estimate by \cite{Bustamante_2021}, $11.76 \pm 0.66 \, \rm km \, s^{-1}$,  which accounts for crowding effects.

We acknowledge that our simulations cannot exactly reproduce the stellar mass content estimated for Leo~I. Nonetheless, given the significant uncertainties in the properties of the systems investigated, we argue that reproducing the final stellar mass within a factor $<2$ is satisfactory for this proof-of-concept study.
In addition, note that the $\Mblack-M_\star$ relation is characterized by a large scatter along the black hole mass axis, typically quantified in $\sim 0.5$ dex (e.g., \citealt{Shankar_2019, Schutte_2019, Reines_Volonteri_2015}). Hence, the Leo~I system did not have to be located precisely on the $\Mblack-M_\star$ relation, and our mass estimates are to be considered within, at least, a factor of $10^{0.5} \approx 3$.

\begin{figure}%
    \centering
    \includegraphics[angle=0,width=0.5\textwidth]{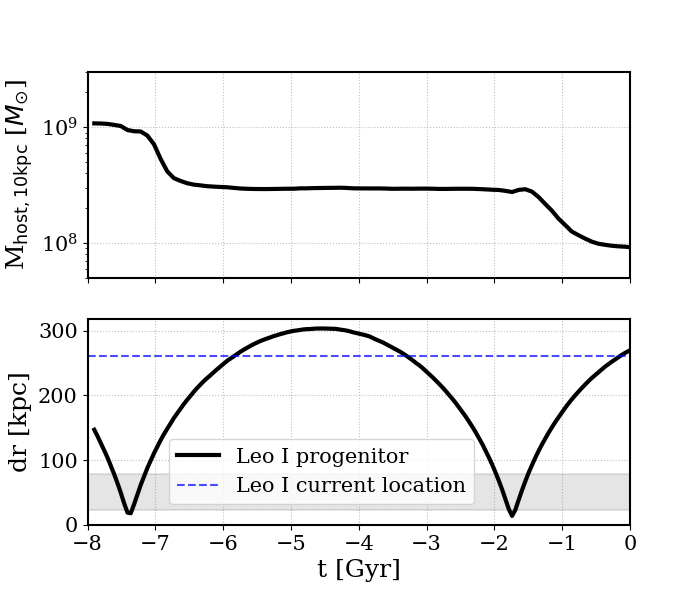}
    \caption{\textbf{Upper panel:} Time evolution of the stellar mass of the progenitor of Leo~I within $10$ kpc, following its evolution of the past $8$ Gyr. The final mass in the nuclear region, within $\sim 1$ kpc, is smaller and compatible with observations. \textbf{Lower panel:} radial absolute distance between the MW center and Leo~I. The horizontal blue line shows the current distance, reproduced at $z=0$. The gray shaded area indicates the range of pericenters estimated with Gaia EDR3 data \citep{Pace_2022}.}
    \label{fig:distance}%
\end{figure}

Figure \ref{fig:distance} shows the evolution of the total stellar mass of Leo~I within $10$ kpc, and its distance to the MW center as a function of time. As Leo~I dramatically changes in mass during our simulation, its characteristic physical dimension shrinks. A radius of $10$ kpc to calculate the enclosed mass allows us to track the stellar mass evolution starting from Leo~I's progenitor, although we acknowledge that some of the mass at large radii may eventually become unbound.

Leo~I's progenitor, initially at $10^9 \Msun$ in stellar mass, becomes a $\sim 3\times 10^8$ object after its first pericenter passage. A mass loss of $70\%$ is remarkably in agreement with our simple analytical model, which predicted $66\%-78\%$ (depending on the velocity dispersion estimate) for the minimum value of the pericenter distance.
After the second passage, the host dark matter halo of Leo~I has been significantly tidally disrupted. The final stellar mass in our simulation, within a radius of $10$ kpc, is $\sim 10^8 \Msun$, with a comparable value for the dark matter mass. After two pericenter passages, the stellar stream of Leo~I is very dispersed, and some of this stellar mass at $r > 1$ kpc is gravitationally unbound. In fact, the final mass in the nuclear region (i.e., within $\sim 1$ kpc) is smaller and compatible with observations.

As shown in Fig. \ref{fig:image}, the two pericenter passages leave behind immense tidal debris. Interestingly, the most recent tidal stream is along our line of sight to Leo~I and is thus challenging to detect.
\cite{Mayer_2008} suggested that a tidal stream of Leo~I could be along our line of sight. This peculiar configuration would explain an observable reversal in its rotational pattern when moving away toward the outskirts of the dSph galaxy. \cite{Mayer_2008} argued that this reversed rotation can be replicated when observing the galaxy from a perspective that aligns closely with its tidal tails. In this scenario, the leading (background) tail gravitationally influences the western region of Leo I, while the trailing (foreground) tail affects the eastern region.

Our N-body is a proof-of-concept and does not prove the existence of the radial tidal tail. More detailed N-body simulations with precise location data are required to scavenge Gaia data and possibly detect the remnant of these giant tidal streams.

\section{Discussion and Conclusions}
\label{sec:conclusions}

Our work was motivated by the dynamical detection of a SMBH of $\sim 3 \times 10^6 \Msun$ at the center of the dSph satellite galaxy Leo~I. If confirmed, this galactic system would be $> 2$ orders of magnitude overmassive with respect to the local $\Mblack-M_\star$ relation; a factor $> 100$ is significantly higher than the typical scatter of $0.5$ dex intrinsic in the relation \citep{Shankar_2019, Schutte_2019, Reines_Volonteri_2015}.

We used a combination of analytical tools and a realistic N-body simulation to investigate a scenario where tidal stripping during one or two pericenter passages with the MW galaxy stripped most of the stellar mass from Leo~I, thus increasing the $\Mblack/M_\star$ ratio.
Our results are summarized as follows:
\begin{itemize}
    \item Our analytical model predicts an average mass loss $\Delta m/m$ for Leo~I's progenitor of $32\%$ to $57\%$ for a single passage at the pericenter, depending on the velocity profile used. The mass loss reaches $66\%$ to $78\%$ if the pericenter distance is the lowest limit constrained by Gaia EDR3 data.
    \item Our N-body simulation reproduces the current position of Leo~I after two passages at the pericenter, with distances compatible with the lowest limit allowed by Gaia EDR3 data.
    \item The mass loss after one passage is $70\%$, in agreement with our simple analytical estimate for low pericenter distances. After two pericenter passages, our simulated system reaches a final mass of $9 \times 10^6 \Msun$ within $1$ kpc, comparable within a factor $<2$ to the estimate in \cite{Mateo_2008}. Considering the large uncertainties estimates of the current stellar and total mass of Leo~I, as well as the intrinsic scatter in the $\Mblack-M_\star$ relation, we argue that this scenario could explain the current state of Leo~I, although its likelihood appears to be low since the requirements are strict (i.e., more than a single, very close pericenter passages).
    \item The two pericenter passages cause giant tidal streams. The most recent tidal stream, the only one still close to the current position of the dSph galaxy, is aligned along our line of sight, making it challenging to detect.
\end{itemize}

More detailed and tailored N-body simulations are required to inform searches in Gaia data of the remnants of these giant tidal streams. Additionally, as noted by \cite{Chang_2013}, the morphology of the progenitor significantly affects the total tidal stripping effect; as such, morphologies that are alternative to the MW-like disk structure we used should be more thoroughly investigated.

If the presence of the SMBH is confirmed, and the tidal stripping hypothesis holds, Leo~I could be a system halfway through the process of becoming a wandering black hole. Based on cosmological simulations, several recent studies show that tidal stripping is a primary mechanism for producing wandering black holes \citep{DiMatteo_2022, Weller_2023}.

In this proof-of-concept study, we have shown that, from a dynamical standpoint, two episodes of extreme tidal stripping of Leo~I may have removed enough stellar mass to displace the system from a location on the $\Mblack-M_\star$ relation to its current, very overmassive locus. 
Although possible in principle, this scenario is disfavored for at least two reasons.

First, the extreme tidal stripping scenario requires two close-by passages with the MW at pericenter distances that are compatible with the minimum allowed by Gaia data.
Second, Leo~I is characterized by a metallicity of $\mathrm{[Fe/H]} = -1.45 \pm 0.01$, which is in line with expectations for a galaxy of its size, given standard stellar mass to stellar metallicity relations (see, e.g., \citealt{Kirby_2013}). If the progenitor of Leo~I were significantly more massive, the average metallicity of its current stellar population would be expected to be higher.

We acknowledge that the metallicity of Leo~I does not favor the extreme tidal stripping hypothesis but argue that an early (i.e., $\sim 7.5$ Gyr ago) pericenter passage could have disrupted the metal enrichment process of the galaxy. The mass-to-metallicity relation is connected to the retention capability of metals in the gravitational potential wells of galaxies \citep{Dekel_Silk_1986, Kirby_2013}. For example, an early pericenter passage could have triggered a starburst \citep{Bekki_2001, Renaud_2014}, which, in turn, removed gas from the galaxy or even suppressed star formation altogether by tidal removal of cold molecular gas \citep{Spilker_2022}. Both these effects could have prevented additional episodes of metal enrichment. We defer the study of these effects to a future investigation via hydrodynamic simulations.

If the SMBH in Leo~I is confirmed, and further simulations and Gaia data do not support the tidal stripping hypothesis, alternative explanations must be sought for forming such an overmassive SMBH. These alternatives could be more cosmological.
Overmassive black holes in dwarfs have now been observed up to $z \sim 0.9$ \citep{Mezcua_2023}.
Additionally, \cite{Pacucci_2023_JWST} found that JWST-detected galaxies at $z=4-7$ host black holes that are overmassive by a factor $10-100\times$ compared to their local counterparts in similar galaxies: the local $M_\bullet-M_\star$ relation is violated at $>3 \sigma$ in the redshift range $z=4-7$.

The low-$z$ overmassive systems could be the remnants of the first black hole seeds that formed in the very early Universe (see, e.g., \citealt{Inayoshi_review_2019, Woods_2019} for a review).
Some of those seed black holes may not have grown enough or may be hosted by galaxies that did not merge and grow significantly. This mechanism could explain the presence of such overmassive black holes in local dwarf galaxies (e.g., \citealt{Reines_2013, Pacucci_MaxMass_2017, Chilingarian_2018, Regan_2023}).

Either way, if confirmed, the SMBH in Leo~I would be an exceptional laboratory for studying the intimate connection between black holes and their host galaxies.

\begin{acknowledgments}
We thank Kung-Yi Su and Lars Hernquist for their helpful discussions on the simulation setup.
We are also grateful to Mike Boylan-Kolchin, Jay Strader, and the anonymous referee for helpful suggestions and comments.
F.P. acknowledges support from a Clay Fellowship administered by the Smithsonian Astrophysical Observatory. Y.N. acknowledges support from the ITC Postdoctoral Fellowship. This work was also supported by the Black Hole Initiative at Harvard University, which is funded by grants from the John Templeton Foundation and the Gordon and Betty Moore Foundation.
\end{acknowledgments}

\noindent \textit{Software:} \texttt{GalPy} \citep{GALPY_2015},  \texttt{Astropy} \citep{Astropy_2013, Astropy_2018, Astropy_2022},  \texttt{Arepo} \citep{Arepo_2020}.

\bibliography{ms}

\begin{thebibliography}{}
\expandafter\ifx\csname natexlab\endcsname\relax\def\natexlab#1{#1}\fi
\providecommand{\url}[1]{\href{#1}{#1}}
\providecommand{\dodoi}[1]{doi:~\href{http://doi.org/#1}{\nolinkurl{#1}}}
\providecommand{\doeprint}[1]{\href{http://ascl.net/#1}{\nolinkurl{http://ascl.net/#1}}}
\providecommand{\doarXiv}[1]{\href{https://arxiv.org/abs/#1}{\nolinkurl{https://arxiv.org/abs/#1}}}

\bibitem[{{Astropy Collaboration} {et~al.}(2013){Astropy Collaboration},
  {Robitaille}, {Tollerud}, {Greenfield}, {Droettboom}, {Bray}, {Aldcroft},
  {Davis}, {Ginsburg}, {Price-Whelan}, {Kerzendorf}, {Conley}, {Crighton},
  {Barbary}, {Muna}, {Ferguson}, {Grollier}, {Parikh}, {Nair}, {Unther},
  {Deil}, {Woillez}, {Conseil}, {Kramer}, {Turner}, {Singer}, {Fox}, {Weaver},
  {Zabalza}, {Edwards}, {Azalee Bostroem}, {Burke}, {Casey}, {Crawford},
  {Dencheva}, {Ely}, {Jenness}, {Labrie}, {Lim}, {Pierfederici}, {Pontzen},
  {Ptak}, {Refsdal}, {Servillat}, \& {Streicher}}]{Astropy_2013}
{Astropy Collaboration}, {Robitaille}, T.~P., {Tollerud}, E.~J., {et~al.} 2013,
  \aap, 558, A33, \dodoi{10.1051/0004-6361/201322068}

\bibitem[{{Astropy Collaboration} {et~al.}(2018){Astropy Collaboration},
  {Price-Whelan}, {Sip{\H{o}}cz}, {G{\"u}nther}, {Lim}, {Crawford}, {Conseil},
  {Shupe}, {Craig}, {Dencheva}, {Ginsburg}, {VanderPlas}, {Bradley},
  {P{\'e}rez-Su{\'a}rez}, {de Val-Borro}, {Aldcroft}, {Cruz}, {Robitaille},
  {Tollerud}, {Ardelean}, {Babej}, {Bach}, {Bachetti}, {Bakanov}, {Bamford},
  {Barentsen}, {Barmby}, {Baumbach}, {Berry}, {Biscani}, {Boquien}, {Bostroem},
  {Bouma}, {Brammer}, {Bray}, {Breytenbach}, {Buddelmeijer}, {Burke},
  {Calderone}, {Cano Rodr{\'\i}guez}, {Cara}, {Cardoso}, {Cheedella}, {Copin},
  {Corrales}, {Crichton}, {D'Avella}, {Deil}, {Depagne}, {Dietrich}, {Donath},
  {Droettboom}, {Earl}, {Erben}, {Fabbro}, {Ferreira}, {Finethy}, {Fox},
  {Garrison}, {Gibbons}, {Goldstein}, {Gommers}, {Greco}, {Greenfield},
  {Groener}, {Grollier}, {Hagen}, {Hirst}, {Homeier}, {Horton}, {Hosseinzadeh},
  {Hu}, {Hunkeler}, {Ivezi{\'c}}, {Jain}, {Jenness}, {Kanarek}, {Kendrew},
  {Kern}, {Kerzendorf}, {Khvalko}, {King}, {Kirkby}, {Kulkarni}, {Kumar},
  {Lee}, {Lenz}, {Littlefair}, {Ma}, {Macleod}, {Mastropietro}, {McCully},
  {Montagnac}, {Morris}, {Mueller}, {Mumford}, {Muna}, {Murphy}, {Nelson},
  {Nguyen}, {Ninan}, {N{\"o}the}, {Ogaz}, {Oh}, {Parejko}, {Parley}, {Pascual},
  {Patil}, {Patil}, {Plunkett}, {Prochaska}, {Rastogi}, {Reddy Janga},
  {Sabater}, {Sakurikar}, {Seifert}, {Sherbert}, {Sherwood-Taylor}, {Shih},
  {Sick}, {Silbiger}, {Singanamalla}, {Singer}, {Sladen}, {Sooley},
  {Sornarajah}, {Streicher}, {Teuben}, {Thomas}, {Tremblay}, {Turner},
  {Terr{\'o}n}, {van Kerkwijk}, {de la Vega}, {Watkins}, {Weaver}, {Whitmore},
  {Woillez}, {Zabalza}, \& {Astropy Contributors}}]{Astropy_2018}
{Astropy Collaboration}, {Price-Whelan}, A.~M., {Sip{\H{o}}cz}, B.~M., {et~al.}
  2018, \aj, 156, 123, \dodoi{10.3847/1538-3881/aabc4f}

\bibitem[{{Astropy Collaboration} {et~al.}(2022){Astropy Collaboration},
  {Price-Whelan}, {Lim}, {Earl}, {Starkman}, {Bradley}, {Shupe}, {Patil},
  {Corrales}, {Brasseur}, {N{\"o}the}, {Donath}, {Tollerud}, {Morris},
  {Ginsburg}, {Vaher}, {Weaver}, {Tocknell}, {Jamieson}, {van Kerkwijk},
  {Robitaille}, {Merry}, {Bachetti}, {G{\"u}nther}, {Aldcroft},
  {Alvarado-Montes}, {Archibald}, {B{\'o}di}, {Bapat}, {Barentsen},
  {Baz{\'a}n}, {Biswas}, {Boquien}, {Burke}, {Cara}, {Cara}, {Conroy},
  {Conseil}, {Craig}, {Cross}, {Cruz}, {D'Eugenio}, {Dencheva}, {Devillepoix},
  {Dietrich}, {Eigenbrot}, {Erben}, {Ferreira}, {Foreman-Mackey}, {Fox},
  {Freij}, {Garg}, {Geda}, {Glattly}, {Gondhalekar}, {Gordon}, {Grant},
  {Greenfield}, {Groener}, {Guest}, {Gurovich}, {Handberg}, {Hart},
  {Hatfield-Dodds}, {Homeier}, {Hosseinzadeh}, {Jenness}, {Jones}, {Joseph},
  {Kalmbach}, {Karamehmetoglu}, {Ka{\l}uszy{\'n}ski}, {Kelley}, {Kern},
  {Kerzendorf}, {Koch}, {Kulumani}, {Lee}, {Ly}, {Ma}, {MacBride}, {Maljaars},
  {Muna}, {Murphy}, {Norman}, {O'Steen}, {Oman}, {Pacifici}, {Pascual},
  {Pascual-Granado}, {Patil}, {Perren}, {Pickering}, {Rastogi}, {Roulston},
  {Ryan}, {Rykoff}, {Sabater}, {Sakurikar}, {Salgado}, {Sanghi}, {Saunders},
  {Savchenko}, {Schwardt}, {Seifert-Eckert}, {Shih}, {Jain}, {Shukla}, {Sick},
  {Simpson}, {Singanamalla}, {Singer}, {Singhal}, {Sinha}, {Sip{\H{o}}cz},
  {Spitler}, {Stansby}, {Streicher}, {{\v{S}}umak}, {Swinbank}, {Taranu},
  {Tewary}, {Tremblay}, {de Val-Borro}, {Van Kooten}, {Vasovi{\'c}}, {Verma},
  {de Miranda Cardoso}, {Williams}, {Wilson}, {Winkel}, {Wood-Vasey}, {Xue},
  {Yoachim}, {Zhang}, {Zonca}, \& {Astropy Project
  Contributors}}]{Astropy_2022}
{Astropy Collaboration}, {Price-Whelan}, A.~M., {Lim}, P.~L., {et~al.} 2022,
  \apj, 935, 167, \dodoi{10.3847/1538-4357/ac7c74}

\bibitem[{{Bekki}(2001)}]{Bekki_2001}
{Bekki}, K. 2001, \apj, 546, 189, \dodoi{10.1086/318231}

\bibitem[{{Bellazzini} {et~al.}(2004){Bellazzini}, {Gennari}, {Ferraro}, \&
  {Sollima}}]{Bellazzini_2004}
{Bellazzini}, M., {Gennari}, N., {Ferraro}, F.~R., \& {Sollima}, A. 2004,
  \mnras, 354, 708, \dodoi{10.1111/j.1365-2966.2004.08226.x}

\bibitem[{{Binney} \& {Tremaine}(2008)}]{Galactic_Dynamics_2008}
{Binney}, J., \& {Tremaine}, S. 2008, {Galactic Dynamics: Second Edition}
  (Princeton University Press)

\bibitem[{{Bovy}(2015)}]{GALPY_2015}
{Bovy}, J. 2015, \apjs, 216, 29, \dodoi{10.1088/0067-0049/216/2/29}

\bibitem[{{Boylan-Kolchin} {et~al.}(2013){Boylan-Kolchin}, {Bullock}, {Sohn},
  {Besla}, \& {van der Marel}}]{BK_2013}
{Boylan-Kolchin}, M., {Bullock}, J.~S., {Sohn}, S.~T., {Besla}, G., \& {van der
  Marel}, R.~P. 2013, \apj, 768, 140, \dodoi{10.1088/0004-637X/768/2/140}

\bibitem[{{Bustamante-Rosell} {et~al.}(2021){Bustamante-Rosell}, {Noyola},
  {Gebhardt}, {Fabricius}, {Mazzalay}, {Thomas}, \&
  {Zeimann}}]{Bustamante_2021}
{Bustamante-Rosell}, M.~J., {Noyola}, E., {Gebhardt}, K., {et~al.} 2021, \apj,
  921, 107, \dodoi{10.3847/1538-4357/ac0c79}

\bibitem[{{Byrd} {et~al.}(1994){Byrd}, {Valtonen}, {McCall}, \&
  {Innanen}}]{Byrd_1994}
{Byrd}, G., {Valtonen}, M., {McCall}, M., \& {Innanen}, K. 1994, \aj, 107,
  2055, \dodoi{10.1086/117015}

\bibitem[{{Caputo} {et~al.}(1999){Caputo}, {Cassisi}, {Castellani}, {Marconi},
  \& {Santolamazza}}]{Caputo_1999}
{Caputo}, F., {Cassisi}, S., {Castellani}, M., {Marconi}, G., \&
  {Santolamazza}, P. 1999, \aj, 117, 2199, \dodoi{10.1086/300838}

\bibitem[{{Chang} {et~al.}(2013){Chang}, {Macci{\`o}}, \& {Kang}}]{Chang_2013}
{Chang}, J., {Macci{\`o}}, A.~V., \& {Kang}, X. 2013, \mnras, 431, 3533,
  \dodoi{10.1093/mnras/stt434}

\bibitem[{{Chilingarian} {et~al.}(2018){Chilingarian}, {Katkov}, {Zolotukhin},
  {Grishin}, {Beletsky}, {Boutsia}, \& {Osip}}]{Chilingarian_2018}
{Chilingarian}, I.~V., {Katkov}, I.~Y., {Zolotukhin}, I.~Y., {et~al.} 2018,
  \apj, 863, 1, \dodoi{10.3847/1538-4357/aad184}

\bibitem[{{Dehnen} {et~al.}(2006){Dehnen}, {McLaughlin}, \&
  {Sachania}}]{Dehnen_2006}
{Dehnen}, W., {McLaughlin}, D.~E., \& {Sachania}, J. 2006, \mnras, 369, 1688,
  \dodoi{10.1111/j.1365-2966.2006.10404.x}

\bibitem[{{Dekel} \& {Silk}(1986)}]{Dekel_Silk_1986}
{Dekel}, A., \& {Silk}, J. 1986, \apj, 303, 39, \dodoi{10.1086/164050}

\bibitem[{{Di Matteo} {et~al.}(2023){Di Matteo}, {Ni}, {Chen}, {Croft}, {Bird},
  {Pacucci}, {Ricarte}, \& {Tremmel}}]{DiMatteo_2022}
{Di Matteo}, T., {Ni}, Y., {Chen}, N., {et~al.} 2023, \mnras, 525, 1479,
  \dodoi{10.1093/mnras/stad2198}

\bibitem[{{Dooley} {et~al.}(2016){Dooley}, {Peter}, {Vogelsberger}, {Zavala},
  \& {Frebel}}]{Dooley_2016}
{Dooley}, G.~A., {Peter}, A. H.~G., {Vogelsberger}, M., {Zavala}, J., \&
  {Frebel}, A. 2016, \mnras, 461, 710, \dodoi{10.1093/mnras/stw1309}

\bibitem[{{Event Horizon Telescope Collaboration} {et~al.}(2022){Event Horizon
  Telescope Collaboration}, {Akiyama}, {Alberdi}, {Alef}, {Algaba}, {Anantua},
  {Asada}, {Azulay}, {Bach}, {Baczko}, {Ball}, {Balokovi{\'c}}, {Barrett},
  {Baub{\"o}ck}, {Benson}, {Bintley}, {Blackburn}, {Blundell}, {Bouman},
  {Bower}, {Boyce}, {Bremer}, {Brinkerink}, {Brissenden}, {Britzen},
  {Broderick}, {Broguiere}, {Bronzwaer}, {Bustamante}, {Byun}, {Carlstrom},
  {Ceccobello}, {Chael}, {Chan}, {Chatterjee}, {Chatterjee}, {Chen}, {Chen},
  {Cheng}, {Cho}, {Christian}, {Conroy}, {Conway}, {Cordes}, {Crawford},
  {Crew}, {Cruz-Osorio}, {Cui}, {Davelaar}, {De Laurentis}, {Deane}, {Dempsey},
  {Desvignes}, {Dexter}, {Dhruv}, {Doeleman}, {Dougal}, {Dzib}, {Eatough},
  {Emami}, {Falcke}, {Farah}, {Fish}, {Fomalont}, {Ford}, {Fraga-Encinas},
  {Freeman}, {Friberg}, {Fromm}, {Fuentes}, {Galison}, {Gammie}, {Garc{\'\i}a},
  {Gentaz}, {Georgiev}, {Goddi}, {Gold}, {G{\'o}mez-Ruiz}, {G{\'o}mez}, {Gu},
  {Gurwell}, {Hada}, {Haggard}, {Haworth}, {Hecht}, {Hesper}, {Heumann}, {Ho},
  {Ho}, {Honma}, {Huang}, {Huang}, {Hughes}, {Ikeda}, {Impellizzeri}, {Inoue},
  {Issaoun}, {James}, {Jannuzi}, {Janssen}, {Jeter}, {Jiang},
  {Jim{\'e}nez-Rosales}, {Johnson}, {Jorstad}, {Joshi}, {Jung}, {Karami},
  {Karuppusamy}, {Kawashima}, {Keating}, {Kettenis}, {Kim}, {Kim}, {Kim},
  {Kim}, {Kino}, {Koay}, {Kocherlakota}, {Kofuji}, {Koch}, {Koyama}, {Kramer},
  {Kramer}, {Krichbaum}, {Kuo}, {La Bella}, {Lauer}, {Lee}, {Lee}, {Leung},
  {Levis}, {Li}, {Lico}, {Lindahl}, {Lindqvist}, {Lisakov}, {Liu}, {Liu},
  {Liuzzo}, {Lo}, {Lobanov}, {Loinard}, {Lonsdale}, {Lu}, {Mao}, {Marchili},
  {Markoff}, {Marrone}, {Marscher}, {Mart{\'\i}-Vidal}, {Matsushita},
  {Matthews}, {Medeiros}, {Menten}, {Michalik}, {Mizuno}, {Mizuno}, {Moran},
  {Moriyama}, {Moscibrodzka}, {M{\"u}ller}, {Mus}, {Musoke}, {Myserlis},
  {Nadolski}, {Nagai}, {Nagar}, {Nakamura}, {Narayan}, {Narayanan},
  {Natarajan}, {Nathanail}, {Fuentes}, {Neilsen}, {Neri}, {Ni}, {Noutsos},
  {Nowak}, {Oh}, {Okino}, {Olivares}, {Ortiz-Le{\'o}n}, {Oyama}, {{\"O}zel},
  {Palumbo}, {Paraschos}, {Park}, {Parsons}, {Patel}, {Pen}, {Pesce},
  {Pi{\'e}tu}, {Plambeck}, {PopStefanija}, {Porth}, {P{\"o}tzl}, {Prather},
  {Preciado-L{\'o}pez}, {Psaltis}, {Pu}, {Ramakrishnan}, {Rao}, {Rawlings},
  {Raymond}, {Rezzolla}, {Ricarte}, {Ripperda}, {Roelofs}, {Rogers}, {Ros},
  {Romero-Ca{\~n}izales}, {Roshanineshat}, {Rottmann}, {Roy}, {Ruiz},
  {Ruszczyk}, {Rygl}, {S{\'a}nchez}, {S{\'a}nchez-Arg{\"u}elles},
  {S{\'a}nchez-Portal}, {Sasada}, {Satapathy}, {Savolainen}, {Schloerb},
  {Schonfeld}, {Schuster}, {Shao}, {Shen}, {Small}, {Sohn}, {SooHoo},
  {Souccar}, {Sun}, {Tazaki}, {Tetarenko}, {Tiede}, {Tilanus}, {Titus},
  {Torne}, {Traianou}, {Trent}, {Trippe}, {Turk}, {van Bemmel}, {van
  Langevelde}, {van Rossum}, {Vos}, {Wagner}, {Ward-Thompson}, {Wardle},
  {Weintroub}, {Wex}, {Wharton}, {Wielgus}, {Wiik}, {Witzel}, {Wondrak},
  {Wong}, {Wu}, {Yamaguchi}, {Yoon}, {Young}, {Young}, {Younsi}, {Yuan},
  {Yuan}, {Zensus}, {Zhang}, {Zhao}, {Zhao}, {Agurto}, {Allardi}, {Amestica},
  {Araneda}, {Arriagada}, {Berghuis}, {Bertarini}, {Berthold}, {Blanchard},
  {Brown}, {C{\'a}rdenas}, {Cantzler}, {Caro}, {Castillo-Dom{\'\i}nguez},
  {Chan}, {Chang}, {Chang}, {Chang}, {Chang}, {Chen}, {Chilson}, {Chuter},
  {Ciechanowicz}, {Colin-Beltran}, {Coulson}, {Crowley}, {Degenaar},
  {Dornbusch}, {Dur{\'a}n}, {Everett}, {Faber}, {Forster}, {Fuchs}, {Gale},
  {Geertsema}, {Gonz{\'a}lez}, {Graham}, {Gueth}, {Halverson}, {Han}, {Han},
  {Hasegawa}, {Hern{\'a}ndez-Rebollar}, {Herrera}, {Herrero-Illana},
  {Heyminck}, {Hirota}, {Hoge}, {Hostler Schimpf}, {Howie}, {Huang}, {Jiang},
  {Jinchi}, {John}, {Kimura}, {Klein}, {Kubo}, {Kuroda}, {Kwon}, {Lacasse},
  {Laing}, {Leitch}, {Li}, {Liu}, {Liu}, {Lin}, {Lu}, {Mac-Auliffe},
  {Martin-Cocher}, {Matulonis}, {Maute}, {Messias}, {Meyer-Zhao},
  {Monta{\~n}a}, {Montenegro-Montes}, {Montgomerie}, {Moreno Nolasco},
  {Muders}, {Nishioka}, {Norton}, {Nystrom}, {Ogawa}, {Olivares}, {Oshiro},
  {P{\'e}rez-Beaupuits}, {Parra}, {Phillips}, {Poirier}, {Pradel}, {Qiu},
  {Raffin}, {Rahlin}, {Ram{\'\i}rez}, {Ressler}, {Reynolds},
  {Rodr{\'\i}guez-Montoya}, {Saez-Madain}, {Santana}, {Shaw}, {Shirkey},
  {Silva}, {Snow}, {Sousa}, {Sridharan}, {Stahm}, {Stark}, {Test},
  {Torstensson}, {Venegas}, {Walther}, {Wei}, {White}, {Wieching}, {Wijnands},
  {Wouterloot}, {Yu}, {Yu (于威)}, \& {Zeballos}}]{EHT_SgrA}
{Event Horizon Telescope Collaboration}, {Akiyama}, K., {Alberdi}, A., {et~al.}
  2022, \apjl, 930, L12, \dodoi{10.3847/2041-8213/ac6674}

\bibitem[{{Fattahi} {et~al.}(2018){Fattahi}, {Navarro}, {Frenk}, {Oman},
  {Sawala}, \& {Schaller}}]{Fattahi_2018}
{Fattahi}, A., {Navarro}, J.~F., {Frenk}, C.~S., {et~al.} 2018, \mnras, 476,
  3816, \dodoi{10.1093/mnras/sty408}

\bibitem[{{Genzel} {et~al.}(2010){Genzel}, {Eisenhauer}, \&
  {Gillessen}}]{Genzel_2010}
{Genzel}, R., {Eisenhauer}, F., \& {Gillessen}, S. 2010, Reviews of Modern
  Physics, 82, 3121, \dodoi{10.1103/RevModPhys.82.3121}

\bibitem[{{Genzel} {et~al.}(2000){Genzel}, {Pichon}, {Eckart}, {Gerhard}, \&
  {Ott}}]{Genzel_2000}
{Genzel}, R., {Pichon}, C., {Eckart}, A., {Gerhard}, O.~E., \& {Ott}, T. 2000,
  \mnras, 317, 348, \dodoi{10.1046/j.1365-8711.2000.03582.x}

\bibitem[{{Genzel} {et~al.}(2003){Genzel}, {Sch{\"o}del}, {Ott}, {Eisenhauer},
  {Hofmann}, {Lehnert}, {Eckart}, {Alexander}, {Sternberg}, {Lenzen},
  {Cl{\'e}net}, {Lacombe}, {Rouan}, {Renzini}, \&
  {Tacconi-Garman}}]{Genzel_2003}
{Genzel}, R., {Sch{\"o}del}, R., {Ott}, T., {et~al.} 2003, \apj, 594, 812,
  \dodoi{10.1086/377127}

\bibitem[{{Ghez} {et~al.}(1998){Ghez}, {Klein}, {Morris}, \&
  {Becklin}}]{Ghez_1998}
{Ghez}, A.~M., {Klein}, B.~L., {Morris}, M., \& {Becklin}, E.~E. 1998, \apj,
  509, 678, \dodoi{10.1086/306528}

\bibitem[{{Ghez} {et~al.}(2005){Ghez}, {Salim}, {Hornstein}, {Tanner}, {Lu},
  {Morris}, {Becklin}, \& {Duch{\^e}ne}}]{Ghez_2005}
{Ghez}, A.~M., {Salim}, S., {Hornstein}, S.~D., {et~al.} 2005, \apj, 620, 744,
  \dodoi{10.1086/427175}

\bibitem[{{Ghez} {et~al.}(2008){Ghez}, {Salim}, {Weinberg}, {Lu}, {Do}, {Dunn},
  {Matthews}, {Morris}, {Yelda}, {Becklin}, {Kremenek}, {Milosavljevic}, \&
  {Naiman}}]{Ghez_2008}
{Ghez}, A.~M., {Salim}, S., {Weinberg}, N.~N., {et~al.} 2008, \apj, 689, 1044,
  \dodoi{10.1086/592738}

\bibitem[{{Hausammann} {et~al.}(2019){Hausammann}, {Revaz}, \&
  {Jablonka}}]{Hausamann_2019}
{Hausammann}, L., {Revaz}, Y., \& {Jablonka}, P. 2019, \aap, 624, A11,
  \dodoi{10.1051/0004-6361/201834871}

\bibitem[{{Henriques} \& {Thomas}(2010)}]{Henriques_2010}
{Henriques}, B. M.~B., \& {Thomas}, P.~A. 2010, \mnras, 403, 768,
  \dodoi{10.1111/j.1365-2966.2009.16151.x}

\bibitem[{{Hernquist}(1990)}]{Hernquist_1990}
{Hernquist}, L. 1990, \apj, 356, 359, \dodoi{10.1086/168845}

\bibitem[{{Inayoshi} {et~al.}(2020){Inayoshi}, {Visbal}, \&
  {Haiman}}]{Inayoshi_review_2019}
{Inayoshi}, K., {Visbal}, E., \& {Haiman}, Z. 2020, \araa, 58, 27,
  \dodoi{10.1146/annurev-astro-120419-014455}

\bibitem[{Kang \& van~den Bosch(2008)}]{Kang_2008}
Kang, X., \& van~den Bosch, F.~C. 2008, The Astrophysical Journal Letters, 676,
  L101

\bibitem[{{Kim} {et~al.}(2014){Kim}, {Abel}, {Agertz}, {Bryan}, {Ceverino},
  {Christensen}, {Conroy}, {Dekel}, {Gnedin}, {Goldbaum}, {Guedes}, {Hahn},
  {Hobbs}, {Hopkins}, {Hummels}, {Iannuzzi}, {Keres}, {Klypin}, {Kravtsov},
  {Krumholz}, {Kuhlen}, {Leitner}, {Madau}, {Mayer}, {Moody}, {Nagamine},
  {Norman}, {Onorbe}, {O'Shea}, {Pillepich}, {Primack}, {Quinn}, {Read},
  {Robertson}, {Rocha}, {Rudd}, {Shen}, {Smith}, {Szalay}, {Teyssier},
  {Thompson}, {Todoroki}, {Turk}, {Wadsley}, {Wise}, {Zolotov}, \& {AGORA
  Collaboration29}}]{AGORA_2014}
{Kim}, J.-h., {Abel}, T., {Agertz}, O., {et~al.} 2014, \apjs, 210, 14,
  \dodoi{10.1088/0067-0049/210/1/14}

\bibitem[{{Kim} {et~al.}(2016){Kim}, {Agertz}, {Teyssier}, {Butler},
  {Ceverino}, {Choi}, {Feldmann}, {Keller}, {Lupi}, {Quinn}, {Revaz},
  {Wallace}, {Gnedin}, {Leitner}, {Shen}, {Smith}, {Thompson}, {Turk}, {Abel},
  {Arraki}, {Benincasa}, {Chakrabarti}, {DeGraf}, {Dekel}, {Goldbaum},
  {Hopkins}, {Hummels}, {Klypin}, {Li}, {Madau}, {Mandelker}, {Mayer},
  {Nagamine}, {Nickerson}, {O'Shea}, {Primack}, {Roca-F{\`a}brega}, {Semenov},
  {Shimizu}, {Simpson}, {Todoroki}, {Wadsley}, {Wise}, \& {AGORA
  Collaboration}}]{Agora_2016}
{Kim}, J.-h., {Agertz}, O., {Teyssier}, R., {et~al.} 2016, \apj, 833, 202,
  \dodoi{10.3847/1538-4357/833/2/202}

\bibitem[{{King}(1962)}]{King_1962}
{King}, I. 1962, \aj, 67, 471, \dodoi{10.1086/108756}

\bibitem[{{Kirby} {et~al.}(2013){Kirby}, {Cohen}, {Guhathakurta}, {Cheng},
  {Bullock}, \& {Gallazzi}}]{Kirby_2013}
{Kirby}, E.~N., {Cohen}, J.~G., {Guhathakurta}, P., {et~al.} 2013, \apj, 779,
  102, \dodoi{10.1088/0004-637X/779/2/102}

\bibitem[{{Kochanek}(1996)}]{Kochanek_1996}
{Kochanek}, C.~S. 1996, \apj, 457, 228, \dodoi{10.1086/176724}

\bibitem[{{{\L}okas} {et~al.}(2008){{\L}okas}, {Klimentowski}, {Kazantzidis},
  \& {Mayer}}]{Mayer_2008}
{{\L}okas}, E.~L., {Klimentowski}, J., {Kazantzidis}, S., \& {Mayer}, L. 2008,
  \mnras, 390, 625, \dodoi{10.1111/j.1365-2966.2008.13661.x}

\bibitem[{{Mateo} {et~al.}(2008){Mateo}, {Olszewski}, \& {Walker}}]{Mateo_2008}
{Mateo}, M., {Olszewski}, E.~W., \& {Walker}, M.~G. 2008, \apj, 675, 201,
  \dodoi{10.1086/522326}

\bibitem[{{Mazzarini} {et~al.}(2020){Mazzarini}, {Just}, {Macci{\`o}}, \&
  {Moetazedian}}]{Mazzarini_2020}
{Mazzarini}, M., {Just}, A., {Macci{\`o}}, A.~V., \& {Moetazedian}, R. 2020,
  \aap, 636, A106, \dodoi{10.1051/0004-6361/202037558}

\bibitem[{{Mezcua} {et~al.}(2023){Mezcua}, {Siudek}, {Suh}, {Valiante},
  {Spinoso}, \& {Bonoli}}]{Mezcua_2023}
{Mezcua}, M., {Siudek}, M., {Suh}, H., {et~al.} 2023, \apjl, 943, L5,
  \dodoi{10.3847/2041-8213/acae25}

\bibitem[{{Navarro} {et~al.}(1997){Navarro}, {Frenk}, \&
  {White}}]{Navarro_1997}
{Navarro}, J.~F., {Frenk}, C.~S., \& {White}, S. D.~M. 1997, \apj, 490, 493,
  \dodoi{10.1086/304888}

\bibitem[{{Pace} {et~al.}(2022){Pace}, {Erkal}, \& {Li}}]{Pace_2022}
{Pace}, A.~B., {Erkal}, D., \& {Li}, T.~S. 2022, \apj, 940, 136,
  \dodoi{10.3847/1538-4357/ac997b}

\bibitem[{{Pacucci} \& {Loeb}(2022)}]{Pacucci_2022_Leo}
{Pacucci}, F., \& {Loeb}, A. 2022, \apjl, 940, L33,
  \dodoi{10.3847/2041-8213/ac9b21}

\bibitem[{{Pacucci} {et~al.}(2017){Pacucci}, {Natarajan}, \&
  {Ferrara}}]{Pacucci_MaxMass_2017}
{Pacucci}, F., {Natarajan}, P., \& {Ferrara}, A. 2017, \apjl, 835, L36,
  \dodoi{10.3847/2041-8213/835/2/L36}

\bibitem[{{Pacucci} {et~al.}(2023){Pacucci}, {Nguyen}, {Carniani}, {Maiolino},
  \& {Fan}}]{Pacucci_2023_JWST}
{Pacucci}, F., {Nguyen}, B., {Carniani}, S., {Maiolino}, R., \& {Fan}, X. 2023,
  arXiv e-prints, arXiv:2308.12331, \dodoi{10.48550/arXiv.2308.12331}

\bibitem[{{Read} {et~al.}(2006){Read}, {Wilkinson}, {Evans}, {Gilmore}, \&
  {Kleyna}}]{Read_2006}
{Read}, J.~I., {Wilkinson}, M.~I., {Evans}, N.~W., {Gilmore}, G., \& {Kleyna},
  J.~T. 2006, \mnras, 366, 429, \dodoi{10.1111/j.1365-2966.2005.09861.x}

\bibitem[{{Regan} {et~al.}(2023){Regan}, {Pacucci}, \&
  {Bustamante-Rosell}}]{Regan_2023}
{Regan}, J.~A., {Pacucci}, F., \& {Bustamante-Rosell}, M.~J. 2023, \mnras, 518,
  5997, \dodoi{10.1093/mnras/stac3463}

\bibitem[{{Reines} {et~al.}(2013){Reines}, {Greene}, \& {Geha}}]{Reines_2013}
{Reines}, A.~E., {Greene}, J.~E., \& {Geha}, M. 2013, \apj, 775, 116,
  \dodoi{10.1088/0004-637X/775/2/116}

\bibitem[{{Reines} \& {Volonteri}(2015)}]{Reines_Volonteri_2015}
{Reines}, A.~E., \& {Volonteri}, M. 2015, \apj, 813, 82,
  \dodoi{10.1088/0004-637X/813/2/82}

\bibitem[{{Renaud} {et~al.}(2014){Renaud}, {Bournaud}, {Kraljic}, \&
  {Duc}}]{Renaud_2014}
{Renaud}, F., {Bournaud}, F., {Kraljic}, K., \& {Duc}, P.~A. 2014, \mnras, 442,
  L33, \dodoi{10.1093/mnrasl/slu050}

\bibitem[{{Sales} {et~al.}(2010){Sales}, {Helmi}, \& {Battaglia}}]{Sales_2010}
{Sales}, L.~V., {Helmi}, A., \& {Battaglia}, G. 2010, Advances in Astronomy,
  2010, 194345, \dodoi{10.1155/2010/194345}

\bibitem[{{Schutte} {et~al.}(2019){Schutte}, {Reines}, \&
  {Greene}}]{Schutte_2019}
{Schutte}, Z., {Reines}, A.~E., \& {Greene}, J.~E. 2019, \apj, 887, 245,
  \dodoi{10.3847/1538-4357/ab35dd}

\bibitem[{{Shankar} {et~al.}(2019){Shankar}, {Bernardi}, {Richardson},
  {Marsden}, {Sheth}, {Allevato}, {Graziani}, {Mezcua}, {Ricci}, {Penny}, {La
  Franca}, \& {Pacucci}}]{Shankar_2019}
{Shankar}, F., {Bernardi}, M., {Richardson}, K., {et~al.} 2019, \mnras, 485,
  1278, \dodoi{10.1093/mnras/stz376}

\bibitem[{{Sohn} {et~al.}(2013){Sohn}, {Besla}, {van der Marel},
  {Boylan-Kolchin}, {Majewski}, \& {Bullock}}]{Sohn_2013}
{Sohn}, S.~T., {Besla}, G., {van der Marel}, R.~P., {et~al.} 2013, \apj, 768,
  139, \dodoi{10.1088/0004-637X/768/2/139}

\bibitem[{{Sohn} {et~al.}(2007){Sohn}, {Majewski}, {Mu{\~n}oz}, {Kunkel},
  {Johnston}, {Ostheimer}, {Guhathakurta}, {Patterson}, {Siegel}, \&
  {Cooper}}]{Sohn_2007}
{Sohn}, S.~T., {Majewski}, S.~R., {Mu{\~n}oz}, R.~R., {et~al.} 2007, \apj, 663,
  960, \dodoi{10.1086/518302}

\bibitem[{{Spilker} {et~al.}(2022){Spilker}, {Suess}, {Setton}, {Bezanson},
  {Feldmann}, {Greene}, {Kriek}, {Lower}, {Narayanan}, \&
  {Verrico}}]{Spilker_2022}
{Spilker}, J.~S., {Suess}, K.~A., {Setton}, D.~J., {et~al.} 2022, \apjl, 936,
  L11, \dodoi{10.3847/2041-8213/ac75ea}

\bibitem[{{Spitzer}(1987)}]{Spitzer_1987}
{Spitzer}, L. 1987, {Dynamical evolution of globular clusters} (Princeton
  University Press)

\bibitem[{{Springel} {et~al.}(2005){Springel}, {Di Matteo}, \&
  {Hernquist}}]{Springel_DiMatteo_2005}
{Springel}, V., {Di Matteo}, T., \& {Hernquist}, L. 2005, \apjl, 620, L79,
  \dodoi{10.1086/428772}

\bibitem[{{Springel} {et~al.}(2001){Springel}, {Yoshida}, \&
  {White}}]{Springel2001NewA....6...79S}
{Springel}, V., {Yoshida}, N., \& {White}, S. D.~M. 2001, \na, 6, 79,
  \dodoi{10.1016/S1384-1076(01)00042-2}

\bibitem[{{von Hoerner}(1957)}]{vonHoerner_1957}
{von Hoerner}, S. 1957, \apj, 125, 451, \dodoi{10.1086/146321}

\bibitem[{{Wang} {et~al.}(2023){Wang}, {Li}, {Shan}, {Xu}, {Yao}, {Jing},
  {Gao}, {Li}, {Xie}, {Zhu}, {Yang}, \& {Chen}}]{Wang_2023}
{Wang}, C., {Li}, R., {Shan}, H., {et~al.} 2023, arXiv e-prints,
  arXiv:2305.13694, \dodoi{10.48550/arXiv.2305.13694}

\bibitem[{{Weinberger} {et~al.}(2020){Weinberger}, {Springel}, \&
  {Pakmor}}]{Arepo_2020}
{Weinberger}, R., {Springel}, V., \& {Pakmor}, R. 2020, \apjs, 248, 32,
  \dodoi{10.3847/1538-4365/ab908c}

\bibitem[{{Weller} {et~al.}(2023){Weller}, {Pacucci}, {Natarajan}, \& {Di
  Matteo}}]{Weller_2023}
{Weller}, E.~J., {Pacucci}, F., {Natarajan}, P., \& {Di Matteo}, T. 2023,
  \mnras, 522, 4963, \dodoi{10.1093/mnras/stad1362}

\bibitem[{{Woods} {et~al.}(2019){Woods}, {Agarwal}, {Bromm}, {Bunker}, {Chen},
  {Chon}, {Ferrara}, {Glover}, {Haemmerl{\'e}}, {Haiman}, {Hartwig}, {Heger},
  {Hirano}, {Hosokawa}, {Inayoshi}, {Klessen}, {Kobayashi}, {Koliopanos},
  {Latif}, {Li}, {Mayer}, {Mezcua}, {Natarajan}, {Pacucci}, {Rees}, {Regan},
  {Sakurai}, {Salvadori}, {Schneider}, {Surace}, {Tanaka}, {Whalen}, \&
  {Yoshida}}]{Woods_2019}
{Woods}, T.~E., {Agarwal}, B., {Bromm}, V., {et~al.} 2019, \pasa, 36, e027,
  \dodoi{10.1017/pasa.2019.14}

\bibitem[{{Zaritsky} {et~al.}(1989){Zaritsky}, {Olszewski}, {Schommer},
  {Peterson}, \& {Aaronson}}]{Zaritsky_1989}
{Zaritsky}, D., {Olszewski}, E.~W., {Schommer}, R.~A., {Peterson}, R.~C., \&
  {Aaronson}, M. 1989, \apj, 345, 759, \dodoi{10.1086/167947}

\end{thebibliography}
\bibliographystyle{aasjournal}



\end{document}